\documentclass[10pt,twocolumn,letterpaper]{article}

\usepackage[applications]{wacv}      
\usepackage{graphicx}
\usepackage{subcaption}
\usepackage{float}

\usepackage{graphicx}
\usepackage{amsmath}
\usepackage{amssymb}
\usepackage{booktabs}
\usepackage{algorithm}
\usepackage{algpseudocode}

\usepackage[pagebackref,breaklinks,colorlinks]{hyperref}

\usepackage[capitalize]{cleveref}
\crefname{section}{Sec.}{Secs.}
\Crefname{section}{Section}{Sections}
\Crefname{table}{Table}{Tables}
\crefname{table}{Tab.}{Tabs.}

\def\wacvPaperID{1634} 
\def\confName{WACV}
\def\confYear{2025}

\begin{document}

\title{Bridging the Diagnostic Divide: Classical Computer Vision and Advanced AI methods for distinguishing ITB and CD through CTE Scans}

\author{Shashwat Gupta\\
IIT Kanpur\\
India\\
{\tt\small guptashashwatme@gmail.com}
\and
L. Gokulnath\\
IIT Kanpur\\
India\\
{\tt\small lgokulnath@gmail.com}
\and
Akshan Aggarwal\\
IIT Kanpur\\
India\\
{\tt\small akshancs2020@gmail.com}
\and
Mahim Naz\\
SGPGI Lucknow\\
India\\
{\tt\small mahimnaz8858@gmail.com}
\and
Rajnikanth Yadav\\
SGPGI Lucknow\\
India\\
{\tt\small rajani@sgpgi.ac.in}
\and
Priyanka Bagade\\
IIT Kanpur\\
India\\
{\tt\small pbagade@iitk.ac.in}
}
\maketitle

\begin{abstract}
Differentiating between Intestinal Tuberculosis (ITB) and Crohn's Disease (CD) poses a significant clinical challenge due to their similar symptoms, clinical presentations, and imaging features. This study leverages Computed Tomography Enterography (CTE) scans, deep learning, and traditional computer vision to address this diagnostic dilemma. A consensus among radiologists from renowned institutions has recognized the visceral to subcutaneous fat (VF/SF) ratio as a surrogate biomarker for differentiating between ITB and CD. Previously done manually, we propose a novel 2D image computer vision algorithm for auto segmenting subcutaneous-fat to automate this ratio calculation, enhancing diagnostic efficiency and objectivity. As benchmark, we compare the results to those obtained using TotalSegmentator tool (a popular deep learning-based software tool that helps in automatic segmentation of anatomical structures from medical imaging data) and also from manual calculations done by the radiologists. We have also demonstrated the performance on 3D CT volume as well by using a slicing method and offered a benchmark comparison of the algorithm using the TotalSegmentator tool. Further, we propose a scoring approach to incorporate the scores from radiological features, such as fat-ratio and Pulmonary TB probability, into a single score for diagnosis. We have also trained ResNet10 model on a dataset of CTE scans with samples of ITB, CD, and normal patients, achieving an accuracy of 75\%. Recognizing the need for interpretability to gain clinical trust, we integrated explainable AI technique GradCAM with the ResNet10, to explain the model's predictions. Since the dataset size was small (total 100 cases), the feature based scoring system is more reliable and trusted by radiologists, as compared to the deep learning model for disease diagnosis. 
\end{abstract}

\section{Introduction}
The prevalence of chronic granulomatous intestinal illnesses, including Crohn's disease (CD), an inflammatory bowel disease subtype, and intestinal tuberculosis (ITB), is rising in South and Southeast Asian countries \cite{3_singh2017pivot}. CD and ITB have similar symptomatology and features in imaging, endoscopy, and histopathology. However, the treatments for these diseases greatly vary from one another. Thus, it is essential to differentiate ITB from CD to initiate the correct medical care. Since ITB is primarily found in undeveloped and impoverished nations, few studies have concentrated on the challenge of distinguishing it from CD \cite{1_kedia2019differentiating,sharma2020differentiating}.  

Clinicians employ a range of diagnostic features derived from radiological tests, microbiological tests, pathology, endoscopic examinations, clinical features, and serologic and immunological tests \cite{4_cheng2013clinical}. However, less than 50\% of patients exhibit the majority of these characteristics, diminishing the accuracy of disease diagnosis \cite{6_kedia2018emergence}. Recently, neutral oral contrast medium has been employed to investigate suspected Intestinal Tuberculosis (ITB) and Crohn's Disease (CD) using computed tomography enterography (CTE) \cite{7_costa2010ct}. This method displays extraintestinal signs and ensures effective bowel distention, facilitating the localization of diseases and evaluation of various bowel wall augmentation patterns \cite{8_paulsen2006ct}. Nevertheless, the rising incidence of CD and ITB has led to clinician burnout due to the labor-intensive nature of analyzing numerous CTE scans. Traditional techniques, such as repeated measurements, fail to predict increasingly complex data patterns. Consequently, deep learning approaches should be explored as they can automatically discern complex correlations between predictive markers and outcomes.

\begin{figure*}
  \centering
   \includegraphics[width=0.7\linewidth]{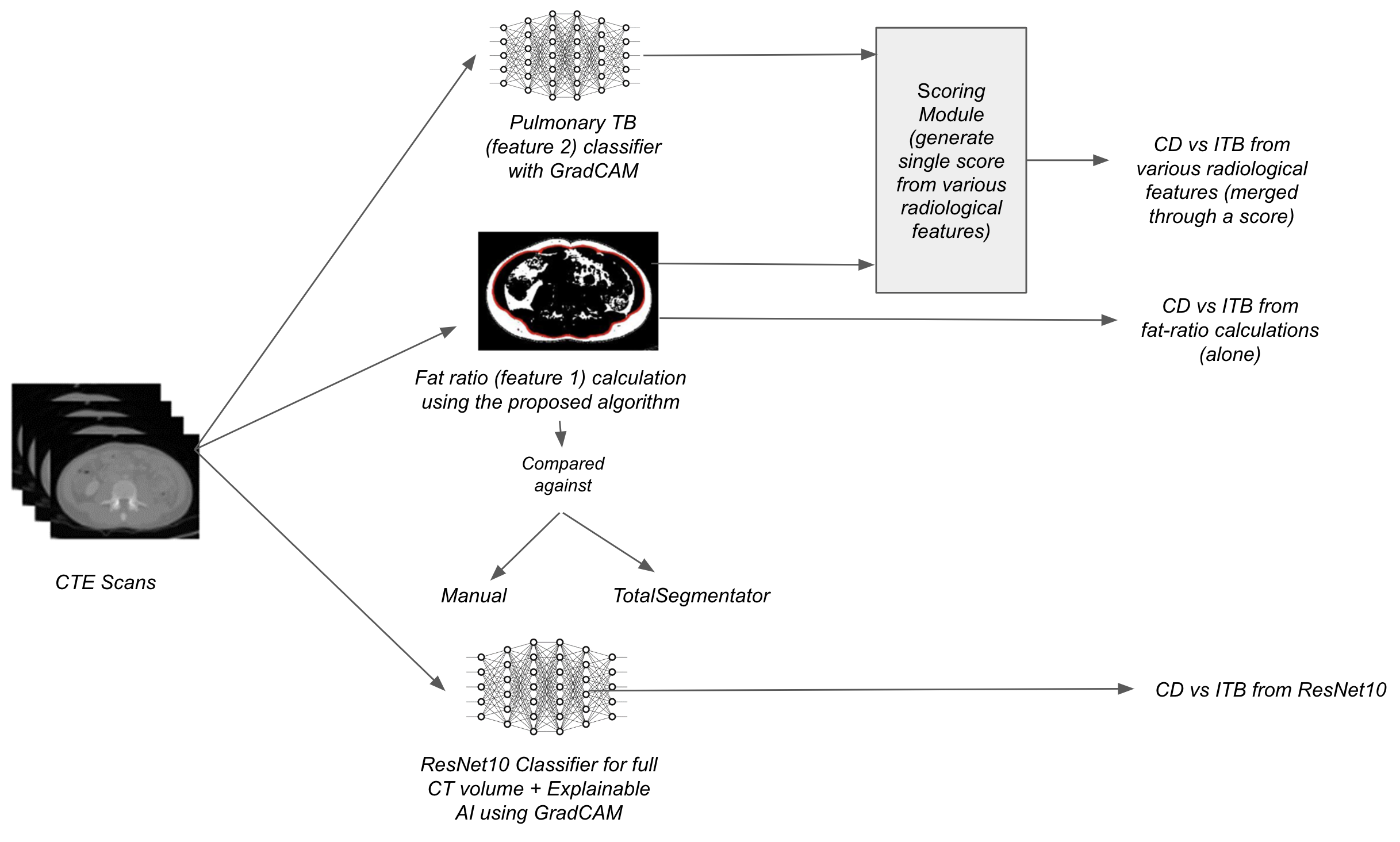}
   \caption{The proposed workflow overview for diagnosing CD vs ITB.}
   \label{fig:methodology}
\end{figure*}

In addition to CTE image analysis, radiologists frequently use the visceral to subcutaneous fat ratio as a surrogate marker to differentiate between Crohn's Disease (CD) and Intestinal Tuberculosis (ITB) \cite{table5_yadav2017development}. However, manually calculating this ratio is labor-intensive. The fat-ratio criterion (Visceral Fat/Subcutaneous Fat $\geq$ 0.63 for CD and otherwise for ITB) is only effective when the ratio significantly deviates from 0.63. Furthermore, radiologists have noted that the presence of pulmonary tuberculosis, although rare due to the infrequency of dual-pathology cases, almost always indicates intestinal TB when intestinal bowel disease is suspected, aiding in the differentiation between Crohn's and intestinal tuberculosis \cite{1_kedia2019differentiating}.

To address these challenges and enhance the current technological tools available to clinicians, we have developed the following techniques:

\begin{enumerate}
    \item We propose a novel classical computer vision algorithm to segment subcutaneous fat and visceral fat based on Hounsfield units (HU) and properties of slices.
    \item We validate and compare the performance of our algorithm with other popular techniques through methods of validation for subcutaneous fat segmentation and area calculation results in 2D (namely, TotalSegmentator (TS) \cite{20_totalsegmentator} and Manual Verification) and in 3D (using TotalSegmentator for generating segmentations). The 3D method (using volume instead of area) generalizes the 2D hypothesis presented by the doctors.
    \item We propose a scoring mechanism to integrate this algorithm with other features for Crohn's Disease and intestinal tuberculosis. This is demonstrated using a pulmonary tuberculosis classifier, explainable using GradCAM and trained on scraped images from Radiopaedia \cite{radiopaedia}.
    \item  We trained ResNet10 model with our custom dataset and deployed the explainable AI technique, GradCAM for explaining the predictions of the classifier.
\end{enumerate}


\begin{algorithm*}
\caption{Algorithm to compute fat ratio}\label{alg:fat_ratio}
\begin{algorithmic}[1]
\Require Image $img$ (grayscale)
\Ensure Total white area and fat ratio
\State $total \gets 0$  \Comment{Initialize total white area}
\For{$x$ in $img$} \Comment{Iterate over rows in the image}
    \For{$y$ in $x$} \Comment{Iterate over pixels in the row}
        \If{$y == 255$} \Comment{Check if pixel is white}
            \State $total \gets total + 1$  \Comment{Increment total white area}
            \State $granular\_degree \gets 0.05$  \Comment{Set granular degree}
        \EndIf
    \EndFor
\EndFor
\State $total\_area \gets total$  \Comment{Store total white area}
\State $subcut\_area \gets$ \text{compute\_subcut\_areas}($img, granular\_degree$)  \Comment{Compute subcutaneous fat area from Algorithm 2}
\State $ratio \gets \frac{total\_area}{subcut\_area} - 1$  \Comment{Compute fat ratio}
\State \textbf{return} $ratio$
\end{algorithmic}
\end{algorithm*}

\begin{algorithm*}
\caption{\texttt{compute\_subcut\_areas(img, granular\_degree)}}
\begin{algorithmic}[1]
\Function{compute\_subcut\_areas}{img, granular\_degree}
    \State $height, width \gets \text{img.shape}[:2]$
    \State $center \gets (\lfloor\frac{width}{2}\rfloor, \lfloor\frac{height}{2}\rfloor)$
    \State $subcut\_area \gets 0$
    
    \For{$\theta \gets 0$ \textbf{to} $360$ \textbf{step} granular\_degree}
        \State $\theta\_rad \gets \theta \times \frac{\pi}{180}$ \Comment{Convert degrees to radians}
        \State $end\_x \gets \text{center}[0] + \cos(\theta\_rad) \times \frac{width}{2}$
        \State $end\_y \gets \text{center}[1] + \sin(\theta\_rad) \times \frac{height}{2}$
        \State $bk, wh \gets \text{find\_last\_point}(img, center, (end\_x, end\_y))$ \Comment{Call to Algorithm 3}
        
        \If{$bk$ \textbf{and} $wh$}
            \State $d1 \gets (\text{wh}[0] - \text{center}[0])^2 + (\text{wh}[1] - \text{center}[1])^2$
            \State $d2 \gets (\text{bk}[0] - \text{center}[0])^2 + (\text{bk}[1] - \text{center}[1])^2$
            \State $subcut\_area \gets subcut\_area + 0.5 \times \left( d1 - d2 \right) \times \text{granular\_degree}$
        \EndIf
    \EndFor
    
    \State \Return $subcut\_area$
\EndFunction
\end{algorithmic}
\end{algorithm*}

\begin{algorithm*}
\caption{\texttt{find\_last\_point(img, start, end)}}
\begin{algorithmic}[1]
\Function{find\_last\_point}{img, start, end}
    \State $last\_black\_point \gets$ None
    \State $last\_white\_point \gets$ None
    \State $answer \gets$ None
    
    \For{$(x, y)$ \textbf{in} \texttt{line\_iter(start, end)}}
        \If{$0 \leq x < \text{img.shape}[1]$ \textbf{and} $0 \leq y < \text{img.shape}[0]$}
            \If{\texttt{img[y, x]} == 0}
                \State $last\_black\_point \gets (x, y)$
            \ElsIf{$\text{last\_black\_point}$ \textbf{and} \texttt{img[y, x]} $\neq 0$}
                \State $answer \gets \text{last\_black\_point}$
            \EndIf
            
            \If{\texttt{img[y, x]} == 255}
                \State $last\_white\_point \gets (x, y)$
            \EndIf
        \EndIf
    \EndFor
    
    \State \textbf{return} $answer$, $last\_white\_point$
\EndFunction
\end{algorithmic}
\end{algorithm*}

\section{Related Work}

Previous research on differentiating Crohn's Disease (CD) and Intestinal Tuberculosis (ITB) primarily involved radiologists manually diagnosing the diseases through CTE images \cite{alpana2019ct,israrahmed2021systematic,table4_zhao2015evaluation,table5_yadav2017development,sharma2016intestinal}. These manual evaluation methods are time-consuming and prone to human errors due to the similar clinical manifestations of both diseases.

To address the limitations of manual CTE image inspection and the reliance on experienced radiologists, researchers have explored machine learning (ML) and deep learning (DL) approaches to improve diagnostic accuracy and efficiency \cite{table3_zeng2022differential,10_enchakalody2020machine,11_chen2022differentiating,table6_zhou2023volumetric,table7_kim2021deep,weng2022differentiation}. Hand-crafted features capturing structural changes within the bowel wall, lumen, and affected areas in CTE images have been used to train various ML/DL models, including Random Forest \cite{10_enchakalody2020machine}, nomogram prediction \cite{table3_zeng2022differential}, Fusion Correlation Neural Network (FCNN) \cite{11_chen2022differentiating}, and Convolutional Neural Network (CNN) \cite{table6_zhou2023volumetric}. While these methods have achieved high prediction accuracy, they still require radiologists to manually identify the affected areas and provide them as input features to the model.

Kim et al. \cite{table7_kim2021deep} used colonoscopy images with a 3D CNN model for disease prediction. However, colonoscopy imaging is invasive and does not cover the entire small bowel, which can be affected by CD. In contrast, this paper utilizes non-invasive CTE scans, which provide a detailed view of the abdomen, allowing radiologists to effectively examine both intestinal and extraintestinal structures. To the best of our knowledge, none of the existing methods fully automate the feature extraction process from CTE scans using deep learning to distinguish between CD and ITB without relying on hand-crafted features.

Additionally, we have developed a classical computer vision algorithm to calculate the visceral to subcutaneous fat ratio, a widely used metric by radiologists for differentiating between CD and ITB \cite{table5_yadav2017development}.To further enhance radiologists' confidence in ML model predictions, researchers have employed explainable AI methods to illustrate the significance of each hand-crafted feature in the model's prediction using Shapley values \cite{weng2022differentiation}. In this paper, our proposed algorithm and TotalSegmentator (TS) generate segmentations that can be used as masks to identify regions of interest. We use GradCAM, an explainable AI technique, to visualize the CTE scan sections considered by the AI model we used for predictions.

\section{Proposed Methodology}\label{sec:methodology}
Figure \ref{fig:methodology} shows our proposed approach to diagnose CD vs ITB. We suggest an algorithm to compute the fat-ratio, a significant radiological feature, from CT slices. The computed fat-ratio is then compared between two methods: TotalSegmentator and manual measurement by radiologists. We used pulmonary TB classifier as one of the features to identify ITB. Then, we incorporated a linear scoring system to enhance our predictions by considering fat-ratio and pulmonary TB probability. In addition, we trained the ResNet10 model on our dataset for diagnosis of these diseases. It allowed us to compare prediction accuracies with deep learning models and feature based methods.

We first extract the slices of interest, particularly at the level of the L4 vertebra in the CTE scan using TotalSegmentator, in the axial view (along the z-axis), and then pass these slices to the algorithm module. Since the visceral and subcutaneous fat values in the L4 region are the highest due to proximity to the infected area, a slice from this region is chosen. The algorithm generates segmentations of subcutaneous fat and utilizes the segmentation and HU values to compute the fat ratio. Although we have used TotalSegmentator to extract the slices, other tools or manual extraction methods can also be employed. 

For comparison, the extracted slices are passed as a volume to TotalSegmentator (TS) to segment subcutaneous and torso fat. Additionally, we pass the full CT scan volume to a pre-trained ResNet10 classifier and compare the results. Segmentations generated by the algorithm and TotalSegmentator explain the regions of interest identified by the model. For ResNet10, we use the GradCAM model for explainability of the input scan. 

We can apply the 2D algorithm to 3D images and the 3D TotalSegmentator (TS) to 2D images by utilizing the following principles: (a) 2D is essentially 3D with one dimension reduced to one. Thus, any 3D algorithm can be applied to a 2D image by treating it as a 3D image with unit dimension along the third axis. (b) 3D can be considered as a stack of 2D slices. Therefore, we can process 3D volumes by dividing them into 2D slices.

The following sections explain each of these contributions in detail. Our code is available at: \small(\url{https://anonymous.4open.science/r/CrohnVsITB-9C5C/}).


\subsection{Subcutaneous Fat Using Classical Computer Vision Approach}
Recent studies indicate that the visceral to subcutaneous fat ratio is a valuable biomarker for distinguishing between Normal, ITB, and Crohn's Disease \cite{table5_yadav2017development}. To reduce the burden on radiologists, we present a classical computer vision algorithm to compute this fat ratio. Additionally, we compare the algorithm's efficiency in terms of memory and time requirements.

\begin{figure*}
    \centering
    \begin{subfigure}[t]{0.24\linewidth}
        \centering
        \includegraphics[width=\linewidth]{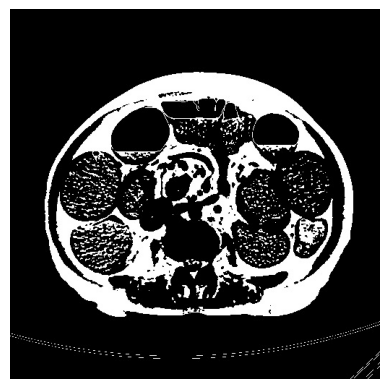}
        \caption{Original CT Scan (binarised based on Hounds Field Unit.}
        \label{fig:a.classical}
    \end{subfigure}
    \hfill
    \begin{subfigure}[t]{0.24\linewidth}
        \centering
        \includegraphics[width=\linewidth]{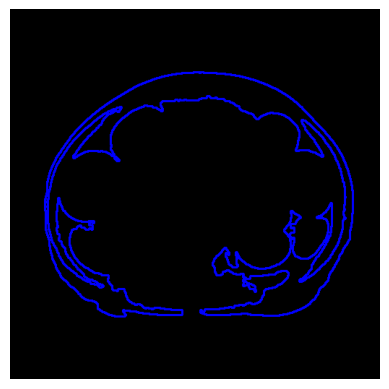}
        \caption{Marking outer boundary (skin) based on contours (erroneous).}
        \label{fig:b.classical}
    \end{subfigure}
    \hfill
    \begin{subfigure}[t]{0.24\linewidth}
        \centering
        \includegraphics[width=\linewidth]{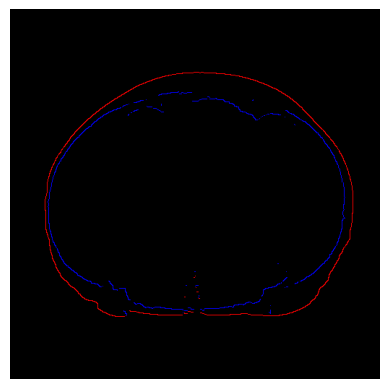}
        \caption{Marking outer and inner boundaries using our algorithm.}
        \label{fig:c.classical}
    \end{subfigure}
    \hfill
    \begin{subfigure}[t]{0.24\linewidth}
        \centering
        \includegraphics[width=\linewidth]{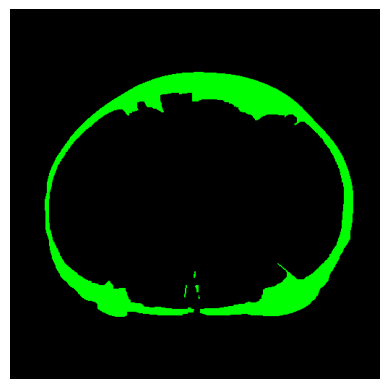}
        \caption{Segmenting the subcutaneous fat using our CV algorithm.}
        \label{fig:d.classical}
    \end{subfigure}
    \caption{Comparison of our algorithm with the classical approach of contours, and segmentation of subcutaneous fat using our algorithm}
    \label{fig:four_images}
\end{figure*}

\subsubsection{Preprocessing}
We extract DICOM (Digital Imaging and Communications in Medicine) slices at the level of the L4 vertebra from the CTE scans along the z-axis using TotalSegmentator. For 2D analysis, we extract the slice at the mean level of L4. For 3D analysis, we extract all slices that contain any portion of the L4 vertebra. A threshold of 0 to -150 Hounsfield Units (HU) (HU range of fat in CT Scans \cite{inproceedings}) is applied to generate a binary mask that includes only fat pixels. Subsequently, we perform one cycle of erosion followed by dilation to remove any additional white lines representing the CT scan machine. This step ensures the proper functioning of our algorithm without significantly altering the fat ratio computation. The boundary and area of the segmented sectors are illustrated in the image below.

\subsubsection{Algorithm}
We compute the boundary between the subcutaneous fat and torso tissues (fat and internal organs) and the skin using the algorithm Algorithm 1. Algorithm 1 finds the boundary point (x,y) from start to end. The boundary point for a given start and end points is defined as the point farthest from center towards the end, which is followed by a continuous sequence of white points and then a continuous sequence of black points. We use this point to mark the boundary of the subcutaneous layer. The algorithm utilises Bresenham's Line Algorithm to compute the points along a given line efficiently. By making incremental decisions based on the slope of the line, Bresenham's algorithm ensures an accurate representation of the line while minimizing the computational complexity of floating-point operations.

\textbf{Algorithm 2} is based on finding area using Polar Form. Using a center as the start and the image border that lies on a ray from center at a given angle, we process points on that line and compute area. The center for our case was approximated to the image center. This assumption worked since we empirically found the image center within the boundary. To generate segmentations, we simply colour the pixels white on the corresponding slice on a black background (of same shape as the extracted slice) as we traverse on the original L4 2D slice from inner boundary of subcutaneous fat to the outer boundary of subcutaneous fat. The runtime of Algorithm 2 is proportional to the volume $\mathit{V}$ of the image, denoted as $O(\mathit{V})$.
Finally, we compute visceral fat to subcutaneous fat ratio in \textbf{Algorithm 3} by substracting 1 from total fat (Algorithm 3) to subcutaneous fat (Algorithm 2) ratio.

\subsection{TotalSegmentator for Segmentations}
To achieve segmentation, we employed the TotalSegmentator tool. Initially, we extracted the L4 region of the CT scans (along the z-axis) using TotalSegmentator. The entire 3D volume was then processed to segment subcutaneous and torso fat across the whole volume. Subsequently, we isolated the L4 slices from the extracted z-axis regions. The volume (or area in 2D) was computed by counting the white pixels in the segmentation mask generated by TotalSegmentator.

\subsection{Classification Model}
\subsubsection{Model Description}
We fine-tune a pre-trained 3D Resnet-10 Model \cite{19_chen2019med3d} originally trained on multiple segmentation datasets of CTE and MRI images. We used the trained feature extraction layers from the pre-trained Resnet10 model and added fully connected layers to perform the classification task. Due to limited data availability, we used the transfer learning approach to improve the accuracy of our model.

\subsubsection{CTE image pre-processing} 
We normalize all the slice volumes to have a mean 0 and standard deviation of 1. We resize all slices to have size 448 $\times$ 448 to match the input size for the pretrained model. Since different CT scans have different number of slices, we use the random slice selection algorithm \cite{18_He2021CovidNet3D} to get equal number of slices for all patients. The algorithms upsamles/downsamples by repeating/deleting slices at randomly chosen locations. We use 600 slices for all patients.

\subsubsection{Model Training} 
The Resnet-10 model is trained for 20 epochs with a step-decay learning rate starting from 0.001. The Cross Entropy loss function is used. The optimizer used is RMSProp. The train-test split was 9:1.
    

\subsubsection{Explanation using GradCAM} 
We use GradCAM \cite{selvaraju2017grad}, a model-agnostic XAI method to validate the observations from the Resnet10 model and possibly point out previously unknown biomarkers. GradCAM is a modification to CAM which does not require any changes to the model. To obtain saliency maps using GradCAM for k-th activation layer (last convolution layer in our case), we compute the gradient ($\frac{\partial y_c}{\partial A^k_{x y z}}$) of class score $y_c$ for class $c$ with respect to feature maps $A^k$ . Gradients are computed during backpropagation. To compute importance ($a_{c}^k$) of each feature map, we utilised the global average pooling over width w, height h and depth d. Then, we perform ReLU activation on the weighted combination of forward activation maps. We use ReLU to activate only the positive pixels which are relevant for our computation. Without ReLU the saliency map is diffused and has noise.

\begin{equation}
a_{c}^k = \frac{1}{Z} \sum^{w}_{x} \sum^{h}_{y} \sum^{d}_{z} \frac{\partial y_c}{\partial A^{k}_{x y z}}
\end{equation}

\begin{equation}
L_{\text{Grad-CAM}}^{c} = \text{ReLU}\left(\sum_{k} a_{k}^{c} A^{k}\right)
\end{equation}

\subsection{Manual Ratio Calculation}
OsiriX Lite, a comprehensive DICOM slice viewer and image processing software, was utilized to annotate medical images in this research. The tool enabled precise annotation and segmentation of regions of interest within the images, allowing for accurate measurement of areas. By leveraging OsiriX Lite’s advanced imaging capabilities, detailed analysis and quantification were performed, significantly contributing to the reliability and accuracy of the research findings. For our purpose, we randomly sampled a subset of 20 CT scans from our entire dataset and calculated the ratio for those using OsiriX Lite software. 

\subsection{Scoring Algorithm with Pulmonary Tuberculosis (PTB) Classifier Example}

We curated a dataset of pulmonary tuberculosis (PTB) cases consisting of approximately 1000 samples, with 500 PTB cases and 500 normal cases. We kept the train:test split ratio as 98:2 and ran for 10 epochs. We trained machine learning models to classify these images, aiming to validate the presence of intestinal TB in the context of suspected intestinal bowel disease by leveraging the probability of PTB derived from the classifier. The classifier's probability output is defined as the softmax of the classifier's logits.

For 3D cases, we extracted the z-coordinates corresponding to the lungs and processed every 10th slice (or another chosen stride length). From these slices, we identified the top three softmax values for TB across all processed slices and averaged them. This average value is referred to as the \textit{P(PTB)}. If this average exceeded 0.5, the case was classified as pulmonary TB, thereby indicating the presence of intestinal tuberculosis in cases where intestinal bowel disease is suspected.

Additionally, we developed a unified scoring system that integrates the contributions of pulmonary TB and Crohn's disease. The scoring formulas are defined as follows:

\begin{equation}
\text{Score}_{\text{Crohn}} = \text{fat\_ratio\_calculated} - 0.63
\end{equation}
\begin{equation}
\text{Score}_{\text{TB}} = a \cdot (0.63 - \text{fat\_ratio\_calculated}) + b \cdot (\text{P(PTB)})
\end{equation}

Here, the coefficients \(a\) and \(b\) can either be set to 1, as done in our initial analysis due to limited data, or can be determined through regression analysis.


\section{Dataset}
\label{sec:Dataset}
The dataset of ITB and CD was acquired in collaboration with a prestigious hospital. It contains Computed Tomography Enterography (CTE) scans, in which the scan is performed after administering a contrast material to the patients to enhance the visibility of the bowel during the scan. Institutional Review Board (IRB) protocol has been followed to shortlist the patient's data. The dataset includes CTE scans data from 36 patients of Crohn's disease, 57 patients of Intestinal Tuberculosis and 7 normal cases. Collaboration with radiologists from the hospital was crucial to identify the key features used to diagnose CD and ITB in CTE scans. Necrotic lymph nodes, and calcified lymph nodes, are the major biomarkers found in patients with ITB, while comb sign, intramural fat, long segment involvement, skip lesions, and fibrofatty proliferation are the biomarkers found in patients with CD.

For pulmonary TB classifier, we scraped the Radiopedia website \cite{radiopaedia} for pulmonary TB and normal chest CT scans.




\section{Results}
\label{results} 
\subsection{2D results for classification and segmentation}
We compared the results obtained from the 2D comparisons of our Computer Vision (CV) based fat ratio algorithm, TotalSegmentator (abbreviated as TS) and the manual calculations. A comparative table based on classification metrics summarising the results can be found in Table \ref{tab:comparison2D}. We also did a segmentation comparison in Table \ref{tab:dice_jaccard} by comparing DICE Scores and Jaccard (Intersection over Union) between Algorithm-and-TotalSegmentator (in case of subcutaneous fat) and Hounsefield Unit (HU)-and-TotalSegmentator (in case of total fat) to offer an idea of degree of overlap between classical and ML approaches.

\begin{table}[h]
\centering
\begin{tabular}{|l|c|c|c|}
\hline
\textbf{} & \textbf{TS} & \textbf{CV Algorithm} & \textbf{Manual} \\ \hline
\textbf{Accuracy} & 0.68 & 0.74 & 0.78 \\ \hline
\textbf{Precision} & 0.58 & 0.67 & 0.76 \\ \hline
\textbf{Recall} & 0.58 & 0.65 & 0.81 \\ \hline
\textbf{Specificity} & 0.74 & 0.80 & 0.75 \\ \hline
\textbf{FDR} & 0.42 & 0.33 & 0.24 \\ \hline
\textbf{NPV} & 0.74 & 0.78 & 0.80 \\ \hline
\textbf{F1} & 0.58 & 0.66 & 0.78 \\ \hline
\textbf{BA} & 0.66 & 0.72 & 0.78 \\ \hline
\textbf{MCC} & 0.32 & 0.45 & 0.56 \\ \hline
\end{tabular}
\caption{Comparison of Classification Metrics for 2D.}
\label{tab:comparison2D}
\end{table}

\begin{table}[h]
\centering
\begin{tabular}{|l|c|c|}
\hline
\textbf{2D} & \textbf{TS and CV Algorithm} & \textbf{TS and HU} \\ \hline
\textbf{DICE} & 0.82±0.23 & 0.78±0.26 \\ \hline
\textbf{Jaccard} & 0.74±0.24 & 0.69±0.26 \\ \hline
\end{tabular}
\caption{Comparison of DICE and Jaccard Indices for 2D.}
\label{tab:dice_jaccard}
\end{table}

Our algorithm was also able to demarcate the boundary of regions, which were not possible with commonly used CV algorithms. For example, figure \ref{fig:four_images} shows a comparison of results of convex-hull, contours and our CV based fat ratio algorithm in segmenting the subcutaneous fat from torso (visceral) fat on an images where fat is white (binarised based on HU values).

\begin{figure*}
  \centering
   \includegraphics[width=0.7\linewidth]{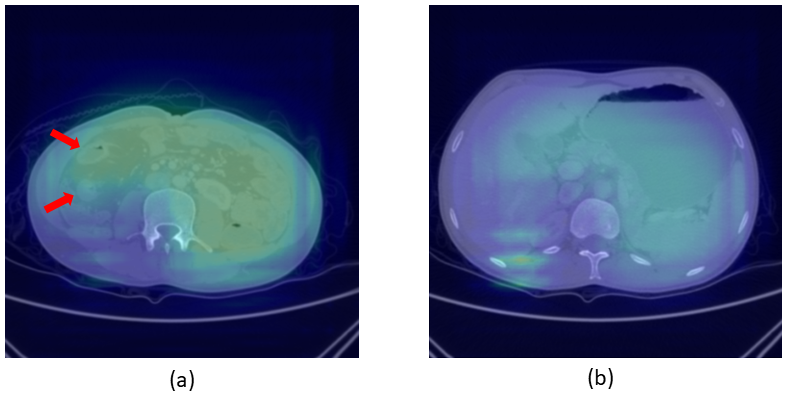}
   \caption{GradCAM results for CD patients (a) Shows the heat map with the comb sign present marked by red arrows, (b) Shows the heatmap of a nearby area without any biomarker for CD.}
   \label{fig:gradcam}
\end{figure*}

The results from manual ratio calculations are regarded as gold standard for ratio values. However, it might still be possible that the ratio is a fair but not absolute indicator for comparison. With respect to the manual annotations, the ratio calculated through the algorithm was found to contain a relative error of 12\%.

\subsection{3D results for classification and segmentation}
Here we compare the results obtained from the 3D comparisons of our Computer Vision (CV) based fat ratio algorithm and TotalSegmentator (TS). A comparative table based on classification metrics summarising the results can be found in Table \ref{tab:comparison3D}. We also did a segmentation comparison by comparing DICE Scores and Jaccard (Intersection over Union) between Algorithm-and-TotalSegmentator (in case of subcutaneous fat) and Hounsefield Unit (HU)-and-TotalSegmentator (in case of total fat) in Table \ref{tab:dice_jaccard_3D} to offer an idea of degree of overlap between classical and ML approaches.

\begin{table}[h]
\centering
\begin{tabular}{|l|c|c|}
\hline
\textbf{3D Comparison} & \textbf{TS} & \textbf{CV Algorithm} \\ \hline
\textbf{Accuracy} & 0.73 & 0.77 \\ \hline
\textbf{Precision (PPV)} & 0.66 & 0.70 \\ \hline
\textbf{Recall (Sensitivity, TPV)} & 0.61 & 0.68 \\ \hline
\textbf{Specificity} & 0.80 & 0.82 \\ \hline
\textbf{FDR} & 0.34 & 0.30 \\ \hline
\textbf{NPV} & 0.77 & 0.80 \\ \hline
\textbf{F1} & 0.63 & 0.69 \\ \hline
\textbf{BA} & 0.71 & 0.75 \\ \hline
\textbf{MCC} & 0.42 & 0.50 \\ \hline
\end{tabular}
\caption{Comparison of Classification Metrics for 3D.}
\label{tab:comparison3D}
\end{table}

\begin{table}[h]
\centering
\begin{tabular}{|l|c|c|}
\hline
\textbf{3D} & \textbf{TS and CV Algorithm} & \textbf{TS and HU} \\ \hline
\textbf{DICE} & 0.82±0.23 & 0.78±0.26 \\ \hline
\textbf{Jaccard} & 0.74±0.24 & 0.69±0.26 \\ \hline
\end{tabular}
\caption{Comparison of DICE and Jaccard Indices for 3D.}
\label{tab:dice_jaccard_3D}
\end{table}

On comparing with the pre-trained deep learning Resnet-10 model for classification, we get a 78\% accuracy on the training set after training (fine-tuning) for 20 epochs and we get a 75\% accuracy on the test set. Figure \ref{fig:gradcam} shows the output of the GradCAM model for a CD patient. The first figure (a) shows the activation map with a comb sign prominently visible (red arrows). The heatmap in this slice is prominently more brightly coloured, showing that the radiologists should focus more on this particular slice. The second figure (b) shows the activation map of a slice from the same patient in a close by region that does not have any prominent biomarker. The heatmap in this slice is prominently green in color, showing that this slice is not an important one. The future study of this work will involve analyzing the additional area shown by the XAI method in CTE scans to find more unnoticed features of these diseases.

\subsection{Performance comparison}
Our CV based fat ratio algorithm was faster on an average by a factor of  3.69 ± 0.79 compared to TS technique. Average runtime (end-to-end excluding file-I/Os and metric calculation) of our algorithm 18.95±7.60 seconds whereas for TotalSegmentator, it was 64.94±15.25 seconds. Both these times exclude the time to find relevant slides along z-axis which was 92.84±24.99 seconds.

\subsection{Pulmonary Classifier for further validation}
On our Radiopedia dataset, we found our model to have 94.9\% accuracy on train set and 90\% accuracy on test set. We further provide the code for GradCam on the models to allow for visualisation of features that contributed to the most for the samples while at inference. The inference is fast and can quickly be used to break the dilemma when fat ratios are close to 0.63. Our dataset had two cases of intestinal Tuberculosis wherein we got the diagnosis of pulmonary tuberculosis from our PTB classifier.



\section{Conclusion and Discussion} 
This paper proposes a novel methodology to distinguish between Crohn's disease and intestinal tuberculosis by deploying feature based method using traditional CV and DL models, as well as compares the performance and efficiency with the existing methods. We collaborated with the radiologists to collect, annotate, and validate the output of our proposed techniques to differentiate between these diseases.

We have developed a classical computer vision algorithm for visceral and subcutaneous fat ratio calculation from the CT scans, with an average relative error of 12\% in our algorithm results with respect to the manual computation outcomes. Moreover, our computer vision algorithm has several advantages over popular ML models or tools and classical computer vision methods. Our CV method has shown faster computations by an average factor of \(3.69 \pm 0.79\) over the TS tool based approach. In addition, it required less data, i.e., by observing the image properties of 10 samples in dataset, we could validate over rest 90 samples, as shown in the results. This required less memory (and does not need any GPU whereas ResNet needed GPU while training) and even performs better than the common tools like ResNet10 and TotalSegmentator. Unlike TotalSegmentator, this approach does not need any annotation. The results of this algorithm demonstrate that while machine learning methods offer great advantages in terms of generalisability, traditional algorithms can outperform the ML models and some of the well-known methods in use-cases such as faster fat-segmentation or in obtaining accurate boundaries. The generalisability and diverse-applicability (eg. ratio calculation, boundary marking, segmentation etc.) of these classical algorithms comes from its modularity (the fact that the each step of the classical algorithm is explainable and tunable) whereas, the ML models are limited in their generalisabilty (only few methods such as changing the last layer exist for these). The algorithm scales with image size (runtime of algorithm is proportional to image volume (so good if small images). Typically ML models need fixed size inputs or should adopt a chunking and selection/sampling strategy.

We have used 100 CTE scans with ResNet10 model, to predict the disease. Since the training data was limited, we used a transfer learning approach, where the ResNet10 model was pre-trained on the CTE image dataset with a 75\% test accuracy. Furthermore, since machine learning models are regarded as black-box, we used explanable AI method called GradCAM to compare the features of model. We further used an explainable AI method, GradCAM, to visualize and understand the areas of the CTE scan images considered by the classifier for predicting the output. Though ML models are black-box, model-agnostic explanable methods such as GradCAM, can be helpful to understand the decision-making ability of models which can be helfpul in designing the cases where models can go wrong and in industries such as healthcare, where explanability is of critical importance. We tried with later versions such as ResNet50 and ResNet34, but improvement in performance was not much significant.

Finally, our linear scoring methodology is applicable to other extracted features, where the confidence (or other metrics similar to confidence) could be linearly weighted to give the total scores; which could then be compared or normalised across a scale.

{\small
\bibliographystyle{ieee_fullname}
\bibliography{egbib}
}

\end{document}